# Watt-level green random laser at 532 nm by SHG of a Yb-doped fiber laser

S. Rota-Rodrigo,[1,*] B. Gouhier,[1] C. Dixneuf,[1,2] L. Antoni-Micollier,[1] G. Guiraud,[2] D. Leandro,[3] M. lopez-amo,[3] N. Traynor,[2] And G. Santarelli[1]

[1]LP2N, Univ. Bordeaux – CNRS – Institut d'Optique Graduate School, F-33400 Talence, France
[2] Azur Light Systems, Pessac, France
[3] Institute of Smart Cities (ISC) and Dept. of Electrical and Electronic Engineering, Public University of Navarra, Navarra, Spain
*Corresponding author: Sergio.rota@institutoptique.fr

We have developed a Watt-level random laser at 532 nm. The laser is based on a 1064 nm random distributed ytterbium-gain assisted fiber laser seed with a 0.35 nm line-width 900mW polarized output power. A study for the optimal length of the random distributed mirror was carried out. An ytterbium-doped fiber master oscillator power amplifier architecture is used to amplify the random seeder laser without additional spectral broadening up to 20 W. By using a periodically poled lithium niobate (PPLN) crystal in a single pass configuration we generate in excess of 1 W random laser at 532 nm by second harmonic generation with an efficiency of 9 %. The green random laser exhibits an instability <1 %, optical signal to noise ratio >70 dB, 0.1 nm linewidth and excellent beam quality.

Random distributed fiber lasers (RDFLs) based on distributed Rayleigh scattering, have been thoroughly investigated due to their high performances and unique features [1]. Where traditional laser schemes are based in resonant cavities for feedback generation, RDFLs use the Rayleigh scattering of a long fiber as distributed mirror, generating a modeless-behavior laser [2,3]. The research in this field has led to the generation of ultra-high power RDFLs from hundreds of Watts [4] to kWs [5,6], narrower linewidth RDFLs up to sub-gigahertz [7], polarized output RDFLs [8-10], tunable RDFLs [11-13] and pulsed generation [14,15]. Gain in RDFLs can be generated from Raman scattering [2,16], by rare earth-doped fibers [12,13] or a hybrid of both [17].

However, to date random lasers in the visible based on second harmonic generation (SHG) of RDFLs have been only reported in [18] with the generation of 110 mw at 654nm in a magnesium periodically poled lithium niobate (MgPPLN) crystal. In this letter, we report for the first time, to the best of our knowledge, a Watt-level visible random laser at 532 nm based on SHG of a RDFL, with a polarized output power in excess of 1W, instability <1 %, optical signal to noise ratio (OSNR) >70 dB and excellent beam quality.

The RDFL is based on a half open-cavity setup assisted by a 3m-long Yb-doped double-clad fiber as gain medium (see Fig. 1). The core and clad radii of the fiber are respectively 10 and 130 µm, and its clad absorption is 4.6 dB/m at 976 nm. The Yb-fiber is forward-pumped through a multi-mode (MM) combiner with a 9W multimode laser diode (LD) at 976nm. Forward pumping has been previously reported as more efficient in Yb-gain assisted RDFL [19]. The wavelength selection is carried out by a high-reflective (99.85%) fiber Bragg grating (FBG) centered at 1064.39 nm and with a 0.57 nm bandwidth. The distributed mirror is based on single-mode fiber (SMF28). Although this fiber operates in multimode regime at 1064 nm, the splice with the 1060 fiber at the output-isolator filters the high order modes. Moreover, the SMF core radius (~9 µm) is comparable to the Yb-fiber one, reducing the losses in the splice. A pump power stripper was used before the SMF fiber in order to remove the residual pump.

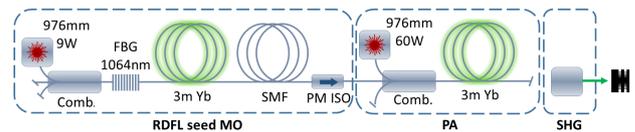

*Fig. 1. Schematic of the 532nm random laser. (RDFL: Random Distributed Fiber Laser, MO: Master Oscillator, PA: Power Amplifier, SHG: Second Harmonic Generation, PM ISO: Polarization Maintaining Isolator)*

To optimize the length of the distributed mirror, we carried out measurements for different SMF lengths from 1.5Km to 3Km. Figure 2 shows the spectra and the output power before the isolator for the different SMF lengths. As expected, Raman scattering becomes significant for longer fibers, starting to be critical for lengths over 3Km. Moreover, the high attenuation of SMF28 at 1064 nm (~1.5 dB/Km) reduce the efficiency, making shorter fibers more attractive. However, the key-point for the distributed mirror length selection was determined by the random laser behavior. RDFL dynamics are less investigated in rare-earth doped-fiber gain assisted systems than in the based on Raman gain. In order to contribute to the understanding of this class of laser, we carried out a consistent study of the RDFL fast dynamics, by using a low-noise fast photodetector with a 6 GHz BW (Discovery Semiconductor, DSC 100S) and an electrical spectrum analyzer (ESA, Agilent N9020A).

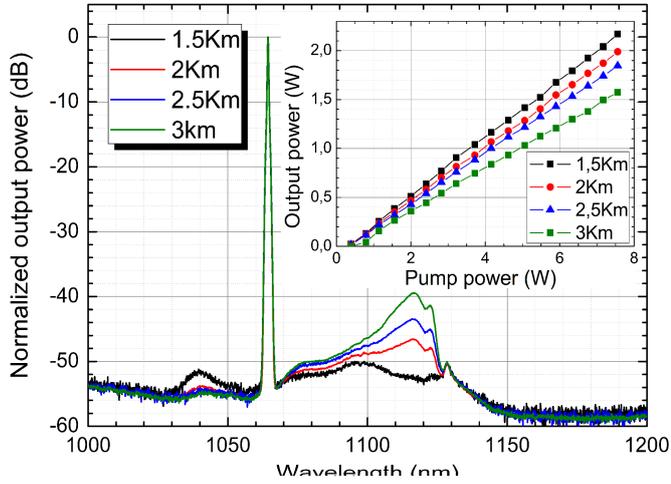

*Fig. 2. RDFL optical spectra at 7.5 W pump for different SMF fiber lengths. Output power Vs pump power for the different fiber lengths (inset)*

In Fig 3, we can see the RDFL radio-frequency power spectrum of the intensity fluctuations for different SMF lengths and pump powers. These results show that under certain conditions, the laser have unstable spurious modes, and is not a pure random distributed feedback laser. This effect has been previously observed in random Raman fiber lasers [2] near the laser threshold. However, in Yb-doped gain assisted RDFL this effect is more pronounced and is observed in a broad power range. We can see that when the pump power is increased, the density of spurious modes starts to decrease. The same effect is achieved by increasing the fiber length. However, the increase of the pump power broadens the laser spectrum probably due to nonlinear wave turbulence effect [11], while this effect is negligible by increasing the fiber length from 1.5 Km to 3 Km.

From these results we can determine that for a pump power of 7.5 W, the optimum SMF length for generating a pure RDFL is around 2.5 Km, with an output power of ~1.9 W and an efficiency of ~25 %, in agreement with [6]. By using a PM isolator and based on the random polarization of the laser, we achieve a PM output power of 0.9 W.

The PM random laser seed was amplified by using the Yb-booster amplified up to 20 W with an efficiency of ~60 %. The booster amplifier is based on a MOPA configuration with a 3m-long Yb-doped PM double-clad fiber 10/125 μm core/clad radii and 4.8 dB absorption at 976 nm. The fiber is pumped with a 60 W thermalized multimode LD at 976 nm through a high-power MM combiner. This amplification stage has been previously used by our group as booster amplifier for single-frequency laser operation at 1064 nm [20].

Fig.4 shows the spectra for different output powers of the booster amplifier, which exhibit a low ASE with an OSNR of 50 dB. It can be seen in the inset that the linewidth of the random seed remains unchanged at the MOPA output.

The second harmonic generation was carried out by using a frequency doubling crystal in a single-pass configuration. The crystal (Covesion, MSHG1064) is a MgO:PPLN with a quasi-phase matching (QPM) 0.21 nm bandwidth. The dimensions of the crystal are 0.5 x 2 x 10 mm and the QPM period is 6.96 μm. Both surfaces are anti-reflection coated at 532 nm and 1064 nm.

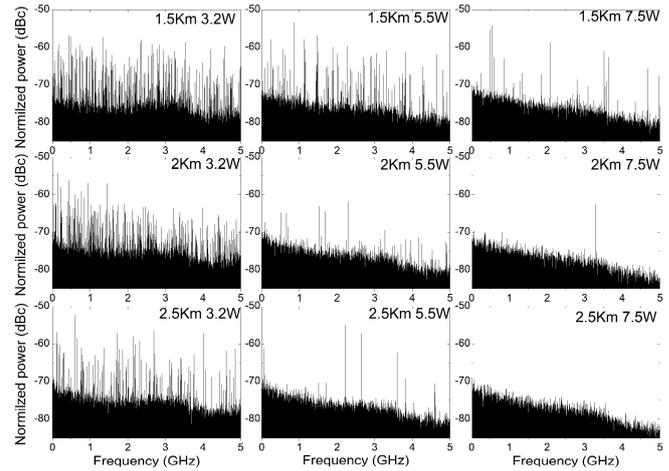

*Fig. 3. Radio-frequency power spectra of the intensity fluctuations of the RDFL seed for different fiber lengths and pump powers measured with ESA. (Resolution BW 100kHz).*

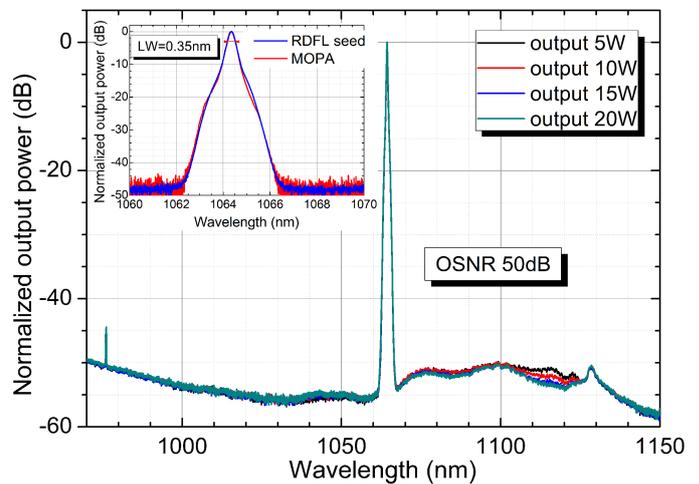

*Fig. 4. Booster amplifier output spectra for different pump powers measured at OSA with resolution of 0.5 nm. (Inset) Short span spectra of the random laser seed for a PM output power of 900 mW (blue) and booster amplifier for an output power of 20 W (red) measured with resolution of 0.01 nm*

The MOPA output beam was focused into the crystal with a waist of 30 μm, by using a 30mm focal lens, corresponding to a focusing parameter ξ=1.88. The optimal waist according to [21] was calculated about 24 μm; however, the use of slightly higher value allows to avoid unwanted effects like thermal lensing or heating [22]. An oven was used for the phase-matching, at around 22.5 °C. The residual IR power at the crystal output was removed by using a dichroic mirror with an extinction ratio of 30 dB.

A characterization of the SHG output power and efficiency of the crystal in function of the MOPA IR output power is shown in Fig. 5 (a). The maximum efficiency of the system was ~9 %, obtaining 1.1 W output power at 532nm. From this point the efficiency starts to decrease. However, we achieved >1.4 W for an input power of 19 W, where the output power starts to exhibit instabilities which could be explained by a high green induced infrared absorption (GRIIRA) effect [23]. Fig. 5 (b) shows the output spectrum of the SHG, with an OSNR higher than 70dB and a linewidth of ~0.1 nm at 532 nm. The latter corresponds to a linewidth of 0.2 nm at 1064 nm which is in accordance with the 0.21 nm phase matching spectral acceptance of the crystal. The efficiency of the system could be improved by matching the linewidth of the laser with the acceptance of the crystal by using a narrower FBG. A shorter crystal would increase the QPM acceptance at the expense of the conversion efficiency.

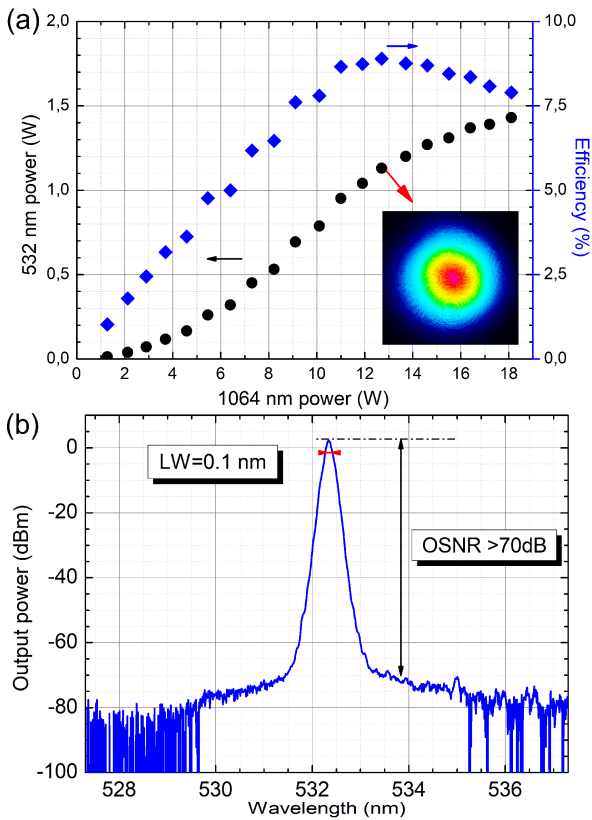

Fig. 5. (a) SHG and efficiency vs pump power. 532nm mode profile at 1.1 W output power (inset). (b) Spectrum of the random 532nm laser at 1.1 W output power measured at OSA with resolution 0.05nm.

The output power level of all the stages was monitored for 1 hour in order to study the power stability of the laser (Fig.6). The random seed presents an instability <5 % (peak-to-peak) which can be explained by the absence of temperature control in the pump LD. However, this power instability is decreased to <0.5 % after the MOPA. This is because the low frequency intensity noise of the RDFL is rejected by the fiber amplifier, in agreement with [24] and the previously reported observations on doped-fiber amplifiers [25,26]. In the SHG process we can notice a slightly increase of the instability up to 1%. From our previous experience with high-power Yb-doped MOPA SF fiber lasers [20,26] and, based on the industrial background of the team [27], we can state that photo-darkening effect in the fiber amplifier is negligible at this power regime and that there is no degradation of the doubling crystal under these operation conditions.

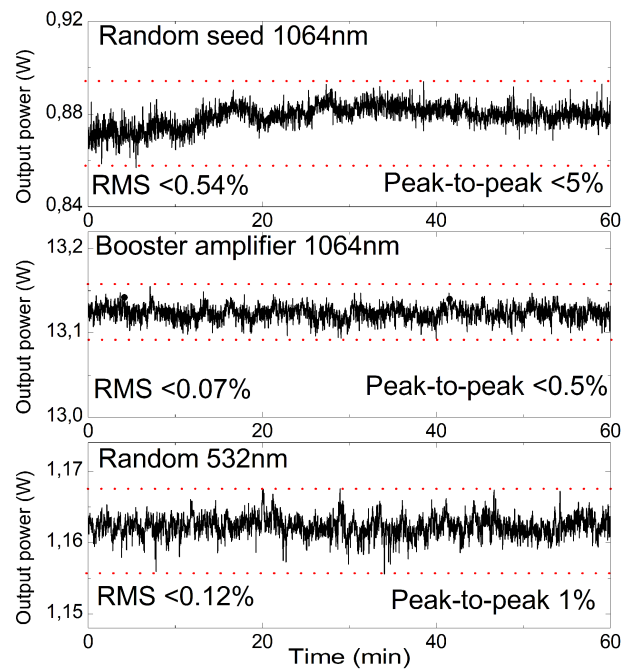

Fig. 6. Stability measurements for random seed (top), booster amplifier (middle) and, second harmonic generation (bottom), along 1 hour. (peak-to-peak fluctuations).

In conclusion, a watt-level green random laser with more than 1 W output power at 532 nm, based on SHG of a RDFL, has been demonstrated. The laser is based on an ytterbium-assisted random distributed fiber laser seed with a 900 mW polarized output power and 0.32 nm linewidth. A study for the optimal length of the random distributed mirror was carried out. An ytterbium-doped fiber master oscillator power amplifier architecture is used to amplify the RDFL output without additional spectral broadening up to 20 W. By using a 10 mm PPLN we generate >1 W random laser at 532nm by second harmonic generation. The green random laser exhibits an instability <1%, optical signal to noise ratio >70 dB, 0.1 nm linewidth and excellent beam quality. This laser can be easily packaged in a robust form for industrial applications.

**Acknowledgment.** We acknowledge the financial support of the cluster of excellence Lasers and Photonics in Aquitane (LAPHIA); Agence Nationale de la Recherche (ANR14 LAB05 0002 01); Conseil Régional d'Aquitaine (2014–IR60309-

00003281). This project has received funding from the European Union's Horizon 2020 research and innovation programme under the Marie Sklodowska-Curie grant agreement No 748839. We also thank the Spanish Government project TEC 2016-76021-C2-1-R, as well as to the AEI/FEDER Funds.

**Link:** https://doi.org/10.1364/OL.43.004284
**Citation:**

S. Rota-Rodrigo, B. Gouhier, C. Dixneuf, L. Antoni-Micollier, G. Guiraud, D. Leandro, M. Lopez-Amo, N. Traynor, and G. Santarelli, "Watt-level green random laser at 532 nm by SHG of a Yb-doped fiber laser," *Opt. Lett.* **43**, 4284-4287 (2018).